\documentclass[twoside,12pt]{article}
\usepackage{wrapfig}
\usepackage{graphicx}
\usepackage{amsmath}
\usepackage{amssymb}
\usepackage{threeparttable}
\usepackage{cite}
\usepackage{xcolor}
\def\single12 {\smallskipamount=3pt plus1pt minus1pt
                  \medskipamount=6pt plus2pt minus2pt
                  \bigskipamount=12pt plus4pt minus4pt
                  \normalbaselineskip=12.8pt plus0pt minus0pt
                  \normallineskip=2pt plus0pt minus0pt
                  \normallineskiplimit=2pt
                  \jot=3pt
                  {\def\smallskip {\vskip\smallskipamount}}
                  {\def\medskip   {\vskip\medskipamount}}
                  {\def\bigskip   {\vskip\bigskipamount}}
                  {\setbox\strutbox=\hbox{\vrule
                    height8.5pt depth3.5pt width 0pt}}
                  \parskip 1pt
                  \normalbaselines}

\def\close10 {\smallskipamount=2pt plus1pt minus1pt
                  \medskipamount=4pt plus2pt minus2pt
                  \bigskipamount=10pt plus4pt minus4pt
                  \normalbaselineskip=12pt plus0pt minus0pt
                  \normallineskip=0pt
                  \normallineskiplimit=0pt
                  \jot=3pt
                  {\def\smallskip {\vskip\smallskipamount}}
                  {\def\medskip   {\vskip\medskipamount}}
                  {\def\bigskip   {\vskip\bigskipamount}}
                  {\setbox\strutbox=\hbox{\vrule
                    height8.5pt depth3.5pt width 0pt}}
                  \parskip 0pt
                  \normalbaselines}

\def\wisk#1{\ifmmode{#1}\else{$#1$}\fi}

\def\deg    {\wisk{^\circ}}

\def\aap{A\&A}
\def\apj{ApJ}
\def\apjs{ApJS}

\def\prd{Phys. Rev. D}         

\def\jcap{JCAP}

\begin{document}

\renewcommand{\bottomfraction}{0.9}
\renewcommand{\topfraction}{0.9}

\begin{center}
{\bf\LARGE Astro2020 APC White Paper}

\vspace{2mm}
{\bf\Large CMB Spectral Distortions: Status and Prospects}

\end{center}

\hspace{17mm} {\bf Primary thematic area:} Cosmology and Fundamental Physics 

\hspace{17mm} {\bf Secondary thematic area:} Galaxy Evolution

\hspace{17mm} {\bf Corresponding author email:} alan.j.kogut@nasa.gov


\vspace{-2mm}

\pagenumbering{roman}  	      
\thispagestyle{empty}

\begin{center}
{\small
A.~Kogut$^{1}$,
M.~H.~Abitbol$^{2}$,
J.~Chluba$^{3}$,
J.~Delabrouille$^{4, 5}$,
D.~Fixsen$^{6}$,
J.~C.~Hill$^{7, 8}$,
S.~P.~Patil$^{9}$,
and 
A.~Rotti$^{3}$
}\\[0mm]
\end{center}


\noindent
{\scriptsize
$^1$ NASA/GSFC, Mail Code: 665, Greenbelt, MD 20771, USA
\\   
%
$^2$ University of Oxford, Denys Wilkinson Building, Keble Road, Oxford, OX1 3RH, UK
\\   
%
$^3$ Jodrell Bank Centre for Astrophysics, School of Physics and Astronomy, The University of Manchester, Manchester M13 9PL, U.K.
\\   
%
$^{4}$ Laboratoire Astroparticule et Cosmologie (APC), CNRS/IN2P3, 10, rue Alice Domon et L\'eonie Duquet, 75205 Paris Cedex 13, France
\\   
%
$^{5}$ D\'epartement d'Astrophysique, CEA Saclay DSM/Irfu, 91191 Gif-sur-Yvette, France 
\\   
%
$^{6}$ Department of Astronomy, University of Maryland, College Park, MD 20742-2421, USA
\\   
%
$^{7}$ Institute for Advanced Study, Princeton, NJ 08540, USA
\\   
%
$^{8}$ Center for Computational Astrophysics, Flatiron Institute, 162 5th Avenue, New York, NY 10010, USA
\\   
%
$^9$ Niels Bohr International Academy and Discovery Center, Blegdamsvej 17, 2100 Copenhagen, Denmark
\\   

}

\vspace{-5mm}
\begin{center}
Submitted in response to the Activity and Project APC call \\
Decadal Survey of Astronomy and Astrophysics 2020
\end{center}


\begin{table}[h]
{
\footnotesize
\begin{center}
\begin{tabular}{l l}
\multicolumn{2}{c}{{\bf Endorsers}} \\
\hline
%
%
Yacine Ali-Ha\"{i}moud	&	Physics Department, New York University\\
Mustafa A. Amin		&	Dept. of Physics and Astronomy, Rice University	\\
Nicola Bartolo		& 	Dipartimento di Fisica e Astronomia, Universita' degli Studi di Padova	\\
Ritoban Basu Thakur	&	California Institute of Technology \\
Ido Ben-Dayan		& 	Ariel University	\\
Boris Bolliet		& 	Jodrell Bank Centre for Astrophysics, University of Manchester	\\
J. Richard Bond		& 	Canadian Institute for Theoretical Astrophysics, University of Toronto	\\
Francois R. Bouchet	&	Institut d'Astrophysique de Paris, CNRS \& Sorbonne Universit\'{e}	\\
Cliff Burgess		&	McMaster University and Perimeter Institute	\\
Carlo Burigana		&	Instituto Nazionale di Aastrofisica (INAF), Istituto di Radioastronomia	\\
Chris Byrnes		&	Department of Physics and Astronomy, University of Sussex	\\
Giovanni Cabass		&	Max-Planck-Institut f\"{u}r Astrophysik	\\
David T. Chuss		&	Department of Physics Villanova University	\\
Sebastien Clesse	& 	Cosmology, Universe and Relativity at Louvain (CURL), University of Louvain	\\
Liang Dai		&	Institute For Advanced Study, Princeton NJ	\\
Vincent Desjacques	&	Physics Department and Asher Space Science Institute	\\
Gianfranco De Zotti	&	Instituto Nazionale di Aastrofisica (INAF)-Osservatorio Astronomico di Padova	\\
Emanuela Dimastrogiovanni  &	School of Physics, The University of New South Wales	\\
Eleonora Di Valentino	&	Jodrell Bank Center for Astrophysics, University of Manchester	\\
Olivier Dore		&	California Institute of Technology		\\
Jo Dunkley		&	Physics Department, Princeton University \\
Ruth Durrer		&	D\'{e}partment de Physique Théorique , Universit\'{e} de Gen\'{e}ve	\\
Cora Dvorkin		&	Department of Physics, Harvard University	\\
H. K. Eriksen		&	Institute of Theoretical Astrophysics, University of Oslo	\\
\hline
\multicolumn{2}{c}{{Continued on next page}} \\
\end{tabular}
\end{center}
}
\end{table}

\clearpage
\begin{table}[h]
{
\footnotesize
\begin{center}
\begin{tabular}{l l}
\multicolumn{2}{c}{{\bf Endorsers (continued)}} \\
\hline
%
%
Tom Essinger-Hileman	&	NASA Goddard Space Flight Center	\\
Matteo Fasiello		&	Institute of Cosmology and Gravitation, University of Portsmouth	\\
Fabio Finelli		&	Instituto Nazionale di Aastrofisica (INAF)	\\
Raphael Flauger		&	Department of Physics, UC San Diego	\\
Juan Garc\'ia-Bellido	&	Instituto de F\'isica Te\'orica, Universidad Aut\'onoma de Madrid	\\
Massimo Gervasi		&	University of Milano Bicocca 	\\
Daniel Grin		&	Haverford College	\\
Diego Herranz		&	Instituto de F\'isica de Cantabria (CSIC-UC) 	\\
Donghui Jeong		&	Department of Astronomy and Astrophysics, The Pennsylvania State University	\\
Bradley R. Johnson	&	Department of Physics, Columbia University	\\
Rishi Khatri		&	Department of Theoretical Physics, Tata Institute of FundamentalResearch	\\
Kazunori Kohri		&	KEK	\\
Kerstin E. Kunze	&	University of Salamanca	\\
John C. Mather		&	NASA Goddard Space Flight Center	\\
Sabino Matarrese	&	Dipartimento di Fisica e Astronomia, Universita' degli Studi di Padova	\\
Joel Meyers		&	Department of Physics, Southern Methodist University	\\
Nareg Mirzatuny		&	Department of Physics and Astronomy, University Of Southern California	\\
Suvodip Mukherjee	&	Institut d'Astrophysique de Paris	\\
Moritz M\"unchmeyer	&	Perimeter Institute for Theoretical Physics	\\
Tomohiro Nakama		&	Institute for Advanced Study, The Hong Kong University of Science and Technology	\\
P.Naselsky		&	Niels Bohr Institute	\\
Federico Nati		&	Department of Physics, University of Milano - Bicocca \\
Elena Orlando		&	Stanford University	\\
Enrico Pajer		&	DAMTP, University of Cambridge	\\
Elena Pierpaoli		&	University of Southern California	\\
Levon Pogosian		&	Department of Physics, Simon Fraser University	\\
Vivian Poulin		& 	CNRS \& Universit\'{e} de Montpellier \\
Andrea Ravenni		&	Jodrell Bank Centre for Astrophysics, The University of Manchester	\\
Christian L. Reichardt	&	School of Physics, University of Melbourne	\\
Mathieu Remazeilles	&	Jodrell Bank Centre for Astrophysics, The University of Manchester	\\
Graca Rocha		&	JPL / California Institute of Technology	\\
Karwan Rostem		&	NASA Goddard Space Flight Center	\\
Jose Alberto Rubi\~no-Martin  &	Instituto de Astrof\'{\i}sica de Canarias	\\
Giorgio Savini		&	Dept. Physics and Astronomy, University College London		\\
Douglas Scott		&	University of British Columbia	\\
Pasquale D. Serpico	&	LAPTh, Univ. Grenoble Alpes	\\
A. A. Starobinsky	&	L. D. Landau Institute for Theoretical Physics	\\
Tarun Souradeep		&	Inter-University Centre for Astronomy and Astrophysics (IUCAA) 	\\
Ravi Subrahmanyan	&	Raman Research Institute, \\
Andrea Tartari		&	Istituto Nazionale di Fisica Nucleare	\\
Tiziana Trombetti	&	CNR, ISMAR Bologna \\
I. K. Wehus		&	Institute of Theoretical Astrophysics, University of Oslo	\\
Siavash Yasini		&	University of Southern California	\\
\hline
\end{tabular}
\end{center}
}
\end{table}

\clearpage
\pagenumbering{arabic}  	      
\setcounter{page}{1}

\large
\noindent
{\bf Executive Summary}

\vspace{3mm}
\normalsize
\single12
\noindent
Departures of the energy spectrum of the cosmic microwave background (CMB)
from a perfect blackbody
probe a fundamental property of the universe -- its thermal history.
Current upper limits, dating back some 25 years,
limit such spectral distortions 
to 50 parts per million 
and provide a foundation 
for the Hot Big Bang model of the early universe.
Modern upgrades to the 1980's-era technology behind these limits
enable three orders of magnitude or greater improvement in sensitivity.
The standard cosmological model 
provides compelling targets at this sensitivity,
spanning cosmic history
from the decay of primordial density perturbations
to the role of baryonic feedback in structure formation.
Fully utilizing this sensitivity
requires concurrent improvements in our understanding
of competing astrophysical foregrounds.
We outline a program using proven technologies
capable of detecting the minimal predicted distortions
even for worst-case foreground scenarios.

\begin{figure}[b]
\vspace{-5mm}
\begin{center}
\begin{tabular}{c}
\includegraphics[height=2.45in]{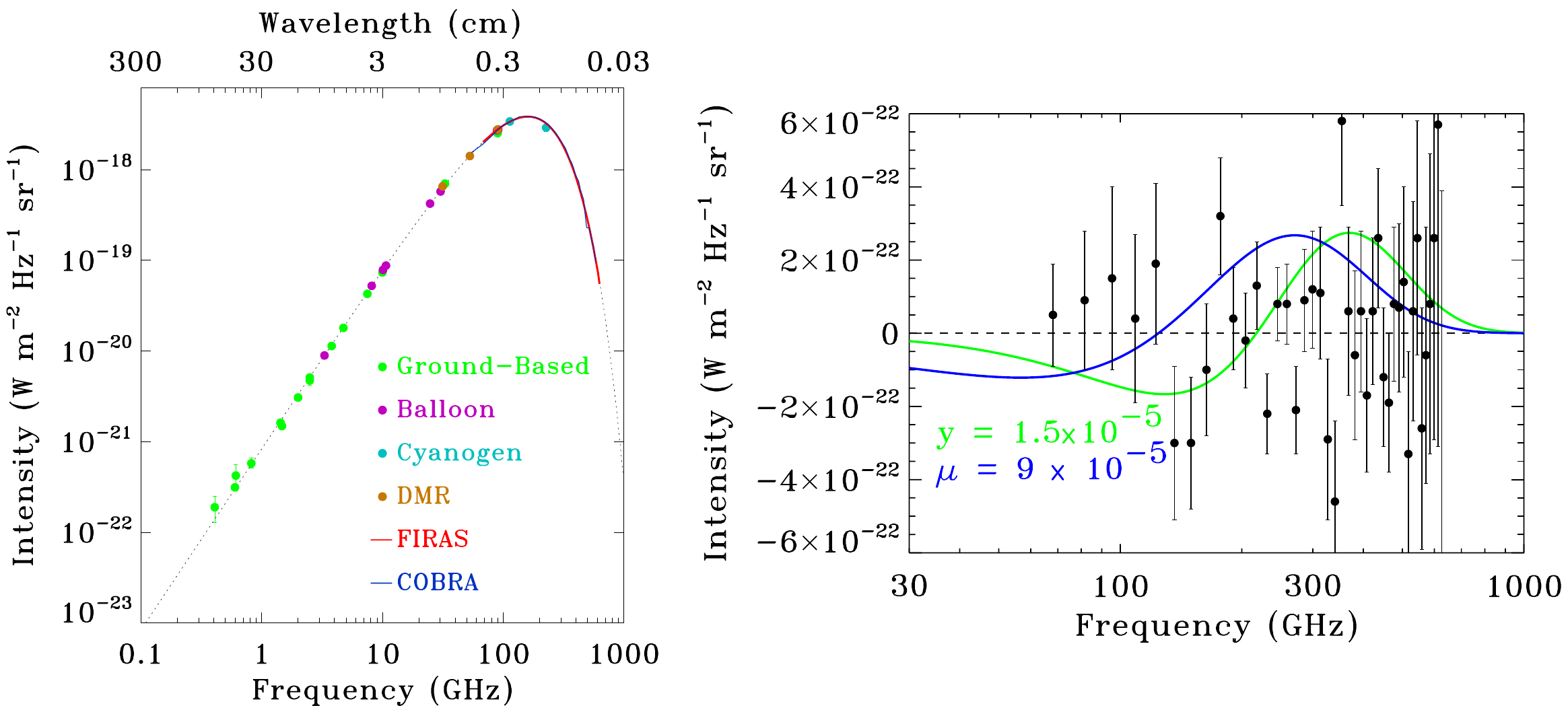}
\end{tabular}
\end{center}
\vspace{-8mm}
\caption[CMB spectrum and spectral distortions]
{\small 
\close10
CMB intensity spectrum and spectral distortions.
(Left) The absolute intensity of the CMB follows 
a blackbody Planck distribution.
(Right) $y$ and $\mu$ spectral distortions
at the FIRAS 95\% confidence upper limit.
\label{spectrum_fig}}
\end{figure}

\large
\vspace{3mm}
\noindent
{\bf 1. Science Goals}

\vspace{3mm}
\normalsize
\single12
\noindent
The cosmic microwave background (CMB)
provides powerful tests for cosmology.
A remnant from the early universe,
today it dominates the sky at millimeter wavelengths.
Its near-perfect blackbody spectrum
provides compelling evidence
for a hot, dense phase at very early times. 
However, deviations from a blackbody
(so-called spectral distortions) are expected and
encode information over the entire thermal history of the universe.
As discussed in \cite{distortion_wp},
spectral distortions result from 
out-of-equilibrium 
energy exchange between matter and radiation.
After energy release into the plasma, 
Compton scattering of CMB photons 
by the electron gas distorts the CMB spectrum
as photons are scattered to higher energies.
Once photon-creating processes become negligible
at redshift $z < 2 \times 10^6$,
the spectrum is unable to evolve back to a (hotter) blackbody,
locking in a distortion whose
amplitude and spectral shape depend on the
epoch, duration, and amplitude of the energy release.
Optically thin scattering 
($z < 10^4$)
creates a Compton $y$-distortion
characterized by the parameter
$y \propto \int n_e (T_e-T_\gamma) dz$
proportional to the integrated electron pressure.
Optically thick scattering
($z > 3 \times 10^5$)
yields the equilibrium Bose-Einstein spectrum,
characterized by the chemical potential
$\mu = 1.4 \Delta E/E$
proportional to the fractional energy release 
relative to the energy in the CMB bath.
Energy release at $10^4 < z < 3 \times 10^5$
produces an intermediate 
spectrum,
%
%
\begin{wrapfigure}[23]{r}{3.3in}
\vspace{-4mm}
\includegraphics[width=3.3in]{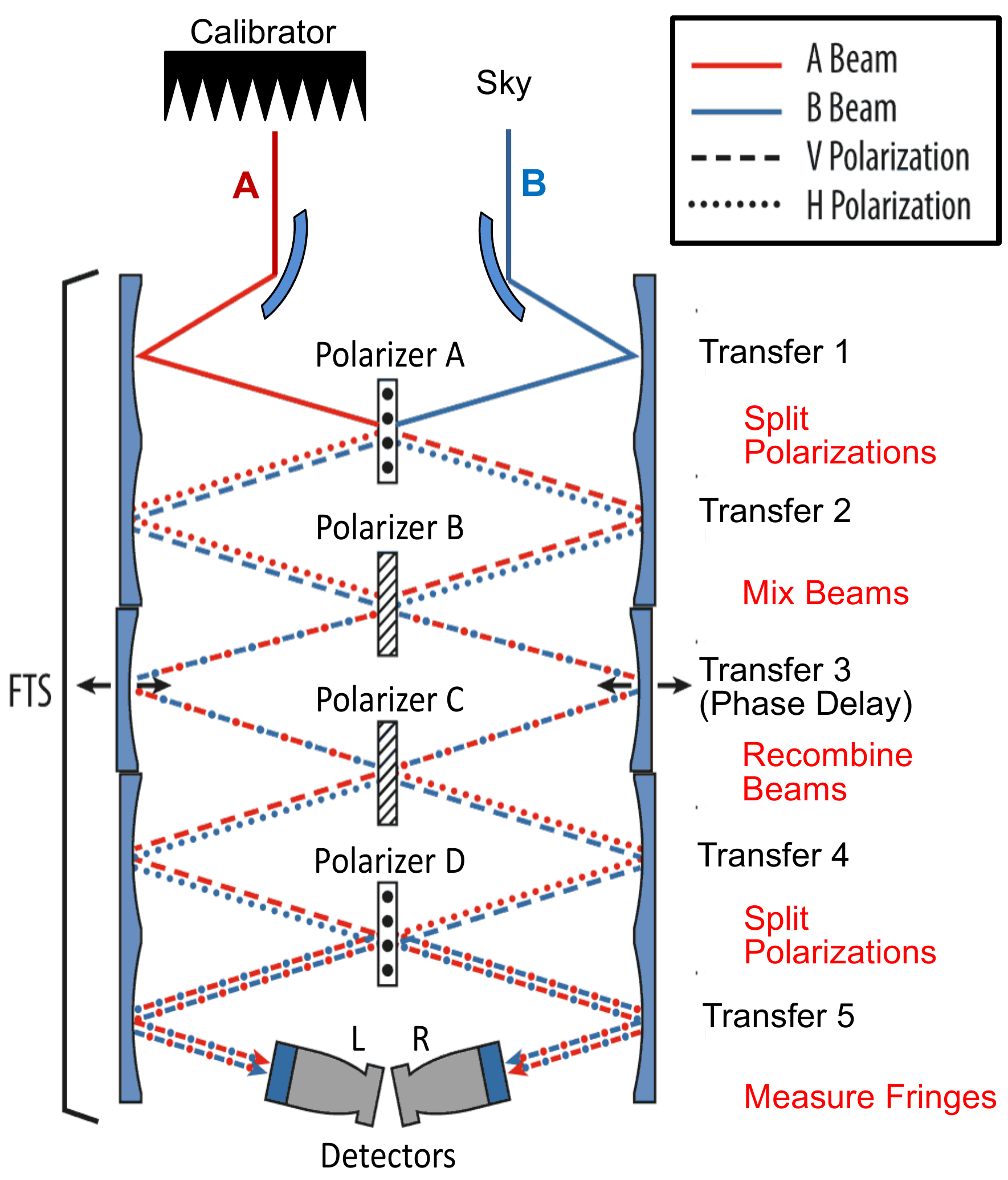}
\vspace{-10mm}
\caption[FTS schematic]
{\small 
\close10
\hspace{-8mm}
Optical signal path
for a fully symmetric Fourier Transform Spectrometer (FTS).}
\label{fts_schematic}
\end{wrapfigure}
%
encoding additional time-dependent 
information
\cite{zeldovich/sunyaev:1969,
illarionov/sunyaev:1974,
chluba/sunyaev:2012,
khatri/sunyaev:2012,
chluba/jeong:2014}. 
Additional rich distortion shapes can be created 
by photon-injection processes 
and interactions with high-energy, non-thermal particles 
\cite{Chluba/greensII, Slatyer2016, Acharya2019}.

Figure~\ref{spectrum_fig} shows 
the CMB blackbody spectrum and spectral distortions.
Current upper limits to spectral distortions
date to the seminal Far Infrared Absolute Spectrophotometer (FIRAS)
measurements in the 1990's.
FIRAS limits spectral distortions to
$|y|~<~15~\times~10^{-6}$
and
$|\mu|~< 9~\times~10^{-5}$
corresponding to fractional distortion
$\Delta I / I < 50$ parts per million
\cite{fixsen/etal:1996}.
Observational progress since FIRAS has been limited.
Measurements in the low-frequency Rayleigh-Jeans tail
from ground-based and balloon platforms
confirmed that the spectrum remains consistent with a blackbody
to 0.1\% at these frequencies,
but did not improve FIRAS constraints on spectral distortions
\cite{bersanelli/etal:1995,
staggs/etal:1996,
gervasi/etal:2008,
fixsen/etal:2011}.

\vspace{3mm}
{\bf Straightforward upgrades to the FIRAS instrument design
would enable breakthrough science.}
FIRAS was not background limited;
its sensitivity was set instead by phonon noise
from its 1.4~K detector.
Modern detectors operating at 0.1 K
have demonstrated phonon noise
well below the intrinsic limit set by photon arrival statistics.
Most of the usable FIRAS data came from a single detector,
whose operational lifetime of 10 months 
ended when the liquid helium ran out.
Combining a modest number of background-limited detectors
with the longer observing times
made possible using mechanical cryocoolers
would improve sensitivity by three orders of magnitude.
{\bf The sky cannot be black at this level:
new measurements provide compelling tests 
of the standard cosmological model
and open a vast discovery space for new physics beyond this model.}
Examples include
spectral distortions induced by the 
decay of long-lived dark matter particles 
(with lifetimes between $t \approx 10^{6}-10^{12}$ s), 
dark matter-standard model particle interactions, 
gravitino decays, axion-photon conversion, 
as well as distortions produced by 
cosmic strings and primordial magnetic fields. 
Spectral distortions are also a powerful probe of 
primordial non-Gaussianity 
and enhanced (or reduced) power in the primordial power spectrum 
at scales far beyond those accessed by CMB anisotropies, 
such as those recently favored by inflationary scenarios 
that produce LIGO mass primordial black holes
\cite{distortion_wp}. 
In addition to this, 
the cosmological standard model 
is expected to produce its own ``floor'' of spectral distortions 
through various mechanisms 
such as the interaction of CMB photons with hot electrons 
during reionization and large-scale structure formation, 
the cosmological recombination process, 
non-equilibrium processes in the pre-recombination hydrogen and helium plasma, 
and the acoustic dissipation of 
small-scale primordial perturbations 
within the standard cosmological scenario
\cite{cabass/etal:2016,Chluba2016LCDM}. 
ASTRO2020 science response 205
\cite{distortion_wp}
summarizes the science from spectral distortions.
In what follows, we lay out a road map 
for observing CMB spectral distortions 
at the sensitivity required to probe this physics.

\clearpage
\large
\noindent
{\bf 2. Measurement Fundamentals}
\vspace{3mm}

\normalsize
\single12
\noindent
Fourier transform spectroscopy is ideally suited 
to search for CMB spectral distortions.
Figure~\ref{fts_schematic} shows the concept.
Two input ports accept light from co-pointed beams on the sky.
A set of five transfer mirror pairs, 
each imaging the previous mirror to the following one, 
shuttles the radiation through a series of polarizing wire grids.
Polarizer A transmits vertical polarization 
and reflects horizontal polarization, 
separating each beam into orthogonal polarization states. 
A second polarizer (B) with wires oriented 45\deg ~relative to grid A 
mixes the polarization states.
A Mirror Transport Mechanism (MTM) moves the central pair of
transfer mirrors
to inject an optical phase delay. 
The phase-delayed beams re-combine (interfere) at Polarizer C. 
Polarizer D (oriented the same as A) 
splits the beams again and routes them to
a set of multi-moded concentrator feed horns.
Each feed contains a pair of identical bolometers,
each sensitive to a single linear polarization 
but mounted at 90\deg ~to each other 
to measure orthogonal polarization states. 
As the MTM sweeps back and forth,
the recombined beams interfere 
to create a fringe amplitude
dependent on the optical phase delay between the two beams.
Let $\vec{E} = E_x \hat{x} + E_y \hat{y}$ 
represent the electric field incident from the sky.
The power at the detectors
as a function of 
frequency $\omega$
and mirror position $z$
may be written
\begin{eqnarray}
P_{Lx} &=& 1/2 ~\smallint \{ ~(E_{Ax}^2+E_{By}^2)+(E_{Ax}^2-E_{By}^2) \cos(4z\omega /c) ~\}{\rm d}\omega~,   \nonumber \\
P_{Ly} &=& 1/2 ~\smallint \{ ~(E_{Ay}^2+E_{Bx}^2)+(E_{Ay}^2-E_{Bx}^2) \cos(4z\omega /c) ~\}{\rm d}\omega~,   \nonumber \\
P_{Rx} &=& 1/2 ~\smallint \{ ~(E_{Ay}^2+E_{Bx}^2)+(E_{Bx}^2-E_{Ay}^2) \cos(4z\omega /c) ~\}{\rm d}\omega~,    \nonumber \\
P_{Ry} &=& 1/2 ~\smallint \{ ~(E_{Ax}^2+E_{By}^2)+(E_{By}^2-E_{Ax}^2) \cos(4z\omega /c) ~\}{\rm d}\omega~,
\label{full_p_eq}
\end{eqnarray}
where L and R refer to the detectors in the left and right concentrators
while A and B refer to the two input beams
(Fig~\ref{fts_schematic}).

We may sample the fringe pattern $P(z)$ measured at each detector
at a set of $N_s$ mirror positions 
to recover the frequency spectrum of the incident radiation.
Let 
$S_\nu$ represent the frequency-dependent sky signal
and
$S_k$ represent the amplitude of the sampled fringe pattern.
The two are related by a Fourier transform, 
\begin{eqnarray}
S_k   &=& \int S_\nu \exp\left( \frac{2\pi i z_k \nu}{c} \right) ~{\rm d}\nu~,	
\qquad
S_\nu 
=
\sum_{k=0}^{N_s-1}{ W_k \, S_k \exp\left( \frac{2\pi i \nu k Z}{c N_s} \right)}~,
\label{syserr_fourier_eq}
\end{eqnarray}
where
$z_k$ is the phase delay for fringe sample $k$,
$W_k$ is the apodization weight,
and
$k$ labels the synthesized frequency channels.
As the mirror moves,
we obtain $N_s$ detector samples
over an optical path length $\pm Z$.
The Fourier transform of the sampled fringe pattern
returns the sky signal at sampled frequencies
$n ~ \times c/(2 Z)$
where
$n = 0, 1, 2, ..., N_s/2$.
The maximum path length (optical stroke)
thus determines the width of the frequency bins
in the synthesized spectra,
while the number of detector samples within each optical stroke
determines the number of frequency bins
and thus the highest sampled frequency.

The noise equivalent power (NEP) of photon noise
in a single linear polarization is determined by
\begin{equation}
{\rm NEP}^2_{\rm photon} = {2A\Omega \over c^2} {(kT)^5\over h^3}
	\int \alpha \epsilon f
	  ~\frac{x^4}{e^x-1} 
	  ~\left( 1 + \frac{\alpha \epsilon f}{e^x-1} \right) ~{\rm d}x~,
\label{mather_12a}
\end{equation}
where
$A$ is the detector area,
$\Omega$ is the detector solid angle,
$\alpha$ is detector absorptivity,
$T$ is the physical temperature of the source,
$\epsilon$ is the emissivity of the source,
and
$f$ is the power transmission through the optics
\cite{mather:1982}.
For a fixed integration time, $\tau$, the detected noise is then simply
\begin{equation}
\delta P = \frac{ {\rm NEP}}{ \sqrt{\tau / 2}}~,
\label{noise_eq}
\end{equation}
where the factor of 2 accounts for the conversion
between the frequency and time domains.
The noise at the detector
may in turn
be referred to the specific intensity on the sky,
\begin{equation}
\delta I_\nu = \frac{ \delta P }
		      { A\Omega ~\Delta \nu ~(\alpha \epsilon f) }~,
\label{i_noise}
\end{equation}
where 
$\Delta \nu$ is the bandwidth of the synthesized frequency channels.
The PIXIE mission concept\cite{kogut/etal:2011} 
presents a worked example
for the sensitivity improvements possible with existing technology.
With etendue 4 cm$^2$ sr
and maximum phase delay of 1 cm,
PIXIE achieves 
spectral sensitivity
for a one-second integration of
$\delta I_\nu = 2.4 \times 10^{-22} 
~{\rm W ~m}^{-2} ~{\rm sr}^{-1} ~{\rm Hz}^{-1} $
within each synthesized frequency channel
of width 15 GHz.

The sensitivity for a background-limited FTS 
depends on the 
collecting area (etendue),
total power absorbed by the detector (optical load),
and the number of synthesized frequency channels (detector sampling).
Several scaling laws are important:

\vspace{2mm}
{\bf Etendue}:
Photon noise increases as the square root of the etendue $A \Omega$
(Eq.~\ref{mather_12a}).
However, since the signal increases {\it linearly} with etendue,
the overall sensitivity {\it improves} as $(A \Omega)^{1/2}$.
For fixed angular resolution on the sky,
the sensitivity improves linearly with the 
diameter of the beam-forming optics;
however,
since etendue must be conserved,
larger collecting area for the beam-forming optics
requires a corresponding increase in the detector area as well.
For fixed collecting area,
the sensitivity thus scales linearly with the angular resolution on the sky.

\vspace{2mm}
{\bf Optical Load}: An FTS is intrinsically broadband. 
The highest synthesized frequency channel depends on the mirror  
throw and detector sampling, but the photon noise depends on 
the total ~ power absorbed ~ from all

%
\begin{wrapfigure}[18]{r}{3.5in}
\vspace{-5mm}
\includegraphics[width=3.5in]{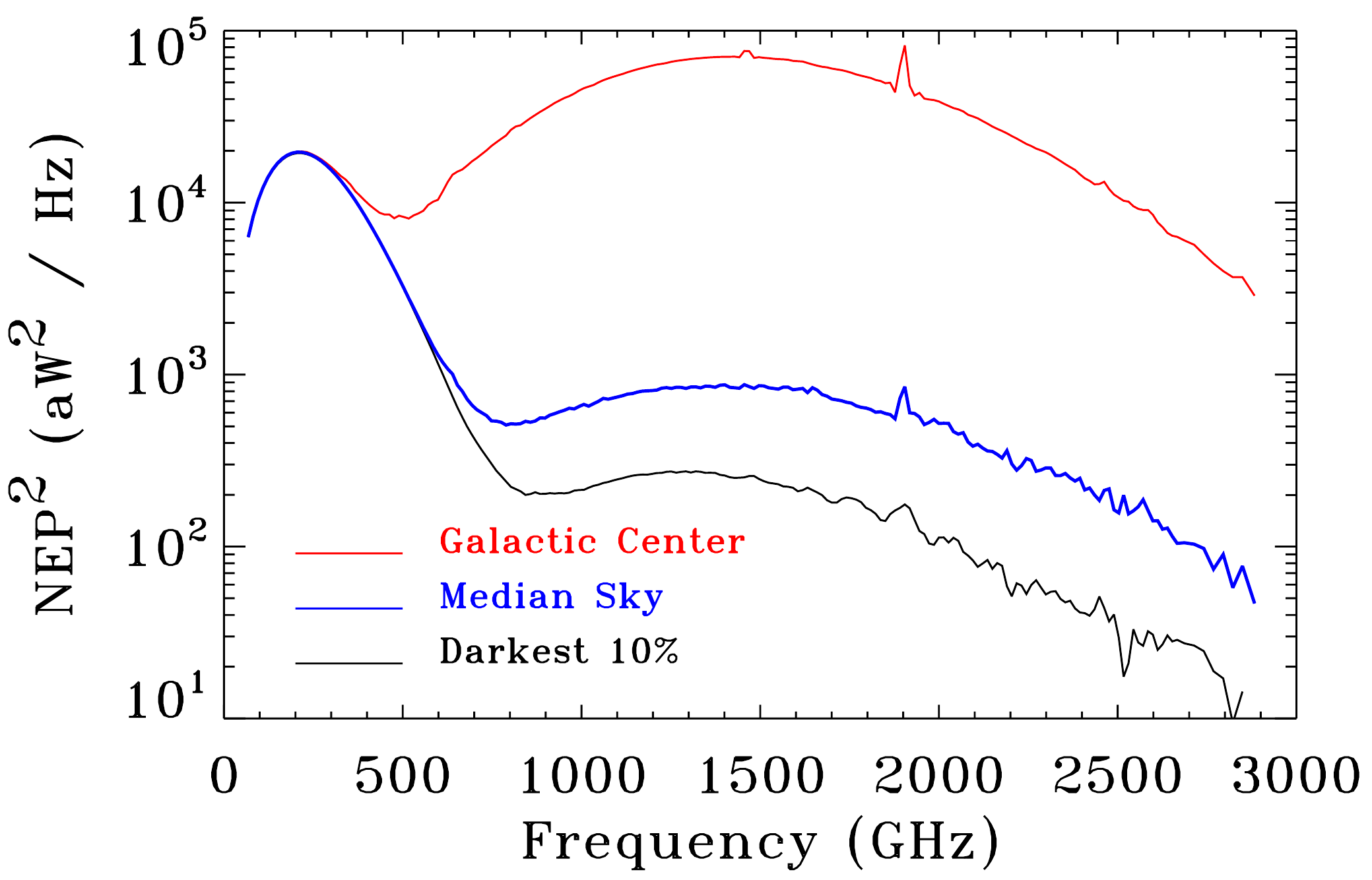}
\vspace{-6mm}
\caption[Photon Noise vs Frequency]
{\small 
\close10
\hspace{-8mm}
The CMB dominates the photon noise budget.
Contributions to the NEP from 
Galactic dust,
zodiacal dust, and the far-IR background
are apparent at frequencies above 600 GHz
but add less than 20\% to the integrated noise.
}
\label{noise_fig}
\end{wrapfigure}
\noindent
frequencies within the instrument passband.
A scattering filter restricts the instrument passband
to limit the noise contribution
and prevent signal aliasing 
from sources at higher frequencies.
At mm wavelengths the sky is dominated by the blackbody CMB,
with lesser contributions
from the far-infrared background,
Galactic dust,
and Solar System zodiacal emission.

Figure~\ref{noise_fig} shows the contribution to the photon noise
from these sources,
as a function of the highest frequency within the instrument passband.
Except for bright regions such as the Galactic center,
extending the instrument passband from
600 GHz to a few THz
increases the noise by less than 20\%.

\clearpage
{\bf Synthesized Channel Width}:
The  noise within each frequency channel
varies inversely with the synthesized channel width $\Delta \nu$
(Eq.~\ref{i_noise}).
Although measurements of line emission
benefit from relatively narrow channels
(to avoid diluting individual lines within broad channels),
continuum spectra such as CMB distortions
benefit from the broadest channel width
consistent with foreground subtraction
($\S$3).
Increasing the mirror throw to generate narrower channels
provides more but noisier channels
within some fixed frequency interval,
degrading the overall sensitivity by
$\Delta \nu^{1/2}$
after co-adding channels.

\begin{figure}[t]
\vspace{-4mm}
\centerline{
\includegraphics[height=2.25in]{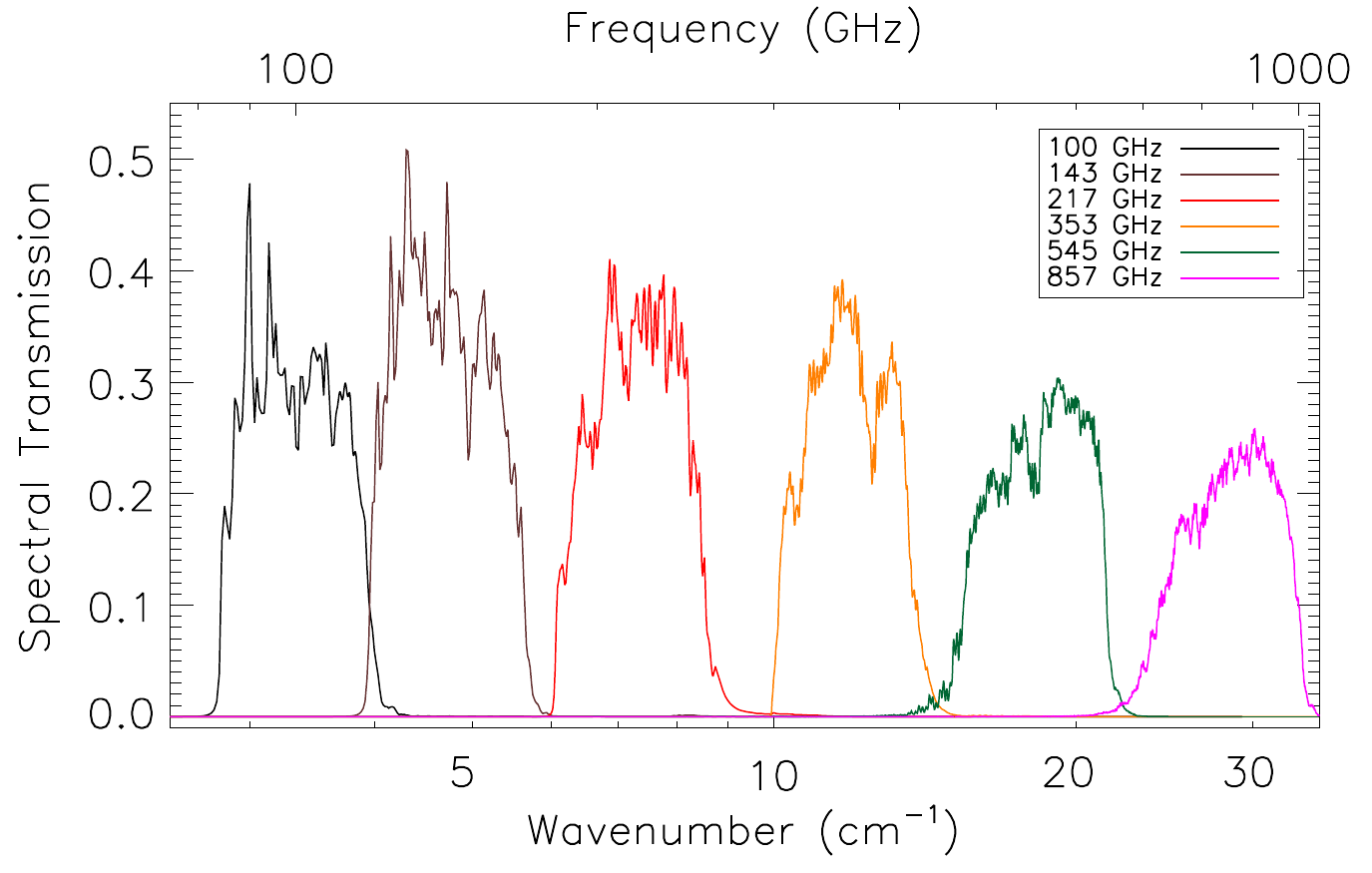}
\includegraphics[height=2.15in]{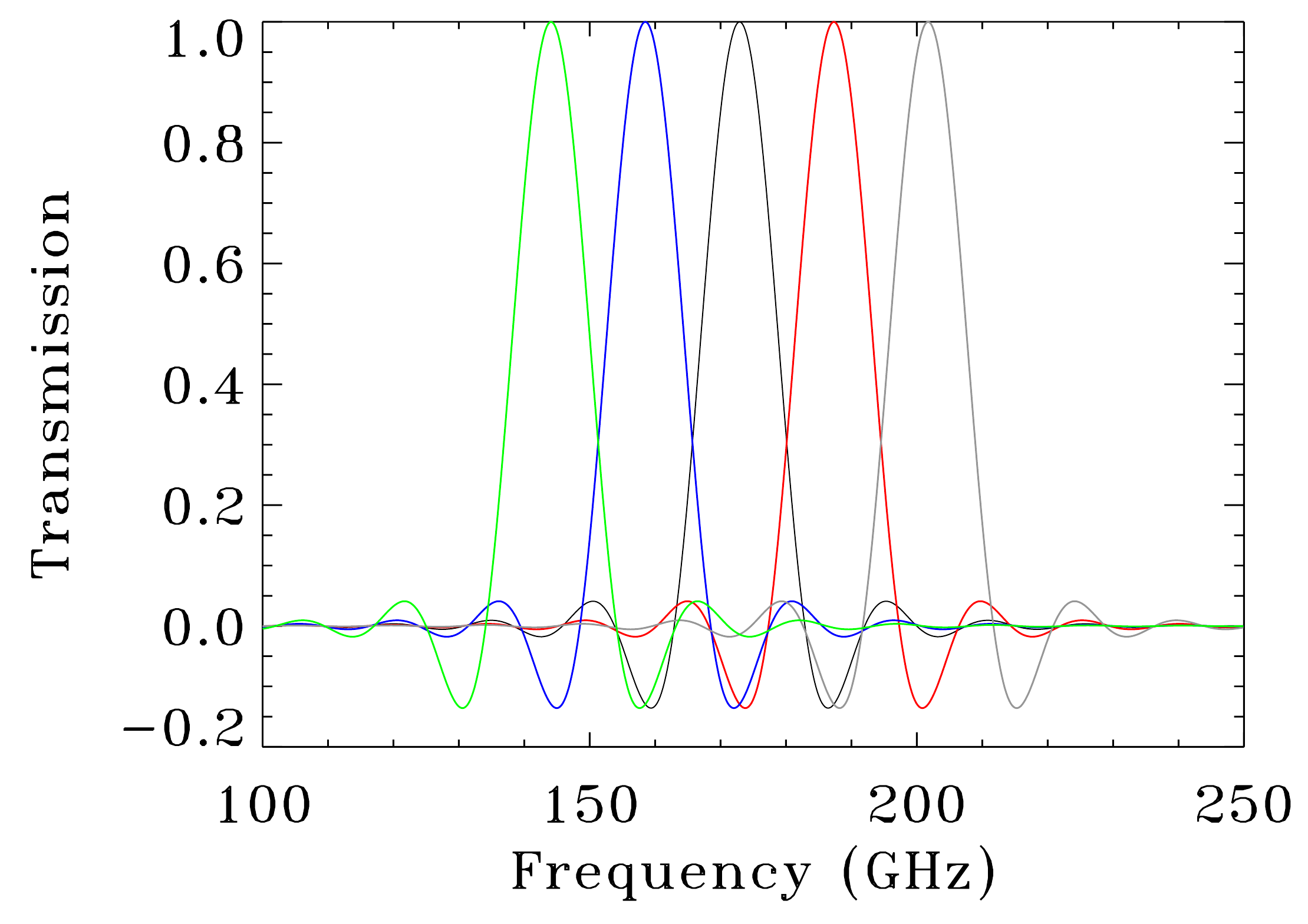}
}
\vspace{-5mm}
\caption[Bandpass Examples]
{\small 
\close10
Bandpass filters for a photometer (left)
depend on device physics and must be measured to high precision.
Synthesized channels for a Fourier transform spectrometer (right)
are determined solely by the fringe sampling and apodization
and can be determined {\it a priori}.}
\label{syserr_bandpass}
\vspace{-4mm}
\end{figure}

\vspace{2mm}
{\bf Synthesized Channel Shape}:
Foregrounds from Galactic and extragalactic sources are brighter
than CMB spectral distortions
($\S$3).
Accurate subtraction of foreground emission
requires measurements at multiple frequencies.
Errors in the frequency response of individual channels
projects foregrounds into the fitted CMB signal,
biasing the estimated spectral distortion.
Fourier transform spectroscopy
offers significant advantages for calibration
and foreground subtraction.
The frequency response for a conventional photometer
is set using physical devices
(quasi-optical filters in the optical path,
lumped-element circuits, etc).
While electromagnetic modeling can predict 
filter performance to few-percent accuracy,
detailed analysis requires supporting measurements
of the as-built filter performance.
The synthesized channels from Fourier transform spectroscopy,
in contrast,
depend only on the sampling and apodization
of the measured fringe pattern
and can be determined {\it a priori}.
The synthesized frequencies may also be set to facilitate
foreground subtraction.
The central frequency of the $k^{th}$ channel
is fixed by the maximum phase delay $Z$,
$\nu_k = k (c/Z)$.
To facilitate subtraction of line emission,
the maximum phase delay $Z$ may be chosen
to be an integer multiple of the wavelength of the
$J=1-0$ CO line,
$Z = M \lambda_{CO}$,
in which case
every $M^{th}$ synthesized channel
is centered on a CO line.

\vspace{2mm}
{\bf Channel Width and Beam Dispersion}:
The synthesized channel width is set 
by the maximum optical phase delay
(Eq.~\ref{syserr_fourier_eq})
and is the same for all channels.
As the phase delay mirror moves,
the change in the optical path length
for a ray along the central axis
differs slightly
from rays at other angles.
As this difference becomes large
compared to the wavelength,
the fringe contrast (signal amplitude) is diminished
when averaged over the beam
while the noise is unaffected.
The resulting signal loss 
at the highest desired frequency $\nu_{\rm max}$
puts a restriction on the
the spectral resolution $\nu / \Delta \nu$
and
the speed of the FTS optics.

\clearpage
\noindent
A simple rule of thumb is
\begin{equation}
f > \frac{ \sqrt{\nu_{\rm max} / \Delta \nu} }{4}
\label{dispersion_eq}
\end{equation}
where $f$ 
(the optical f-number)
is the ratio of diameter to focal length for the FTS transfer mirrors.
Since the etendue must be conserved throughout the entire optical system,
increasing $f$ to obtain higher spectral resolution
at the maximum frequency
requires a corresponding increase in the transfer mirror diameter
and mirror-to-mirror spacing
(Figure~\ref{fts_schematic}).
When the size of the FTS 
exceeds the size of the fore-optics
coupling the FTS to the sky,
it then drives the size and cost of the observatory.

\vspace{3mm}
\large
\noindent
{\bf 3. Foreground Subtraction}

\begin{figure}[b]
\begin{center}
\begin{tabular}{c}
\includegraphics[height=2.9in]{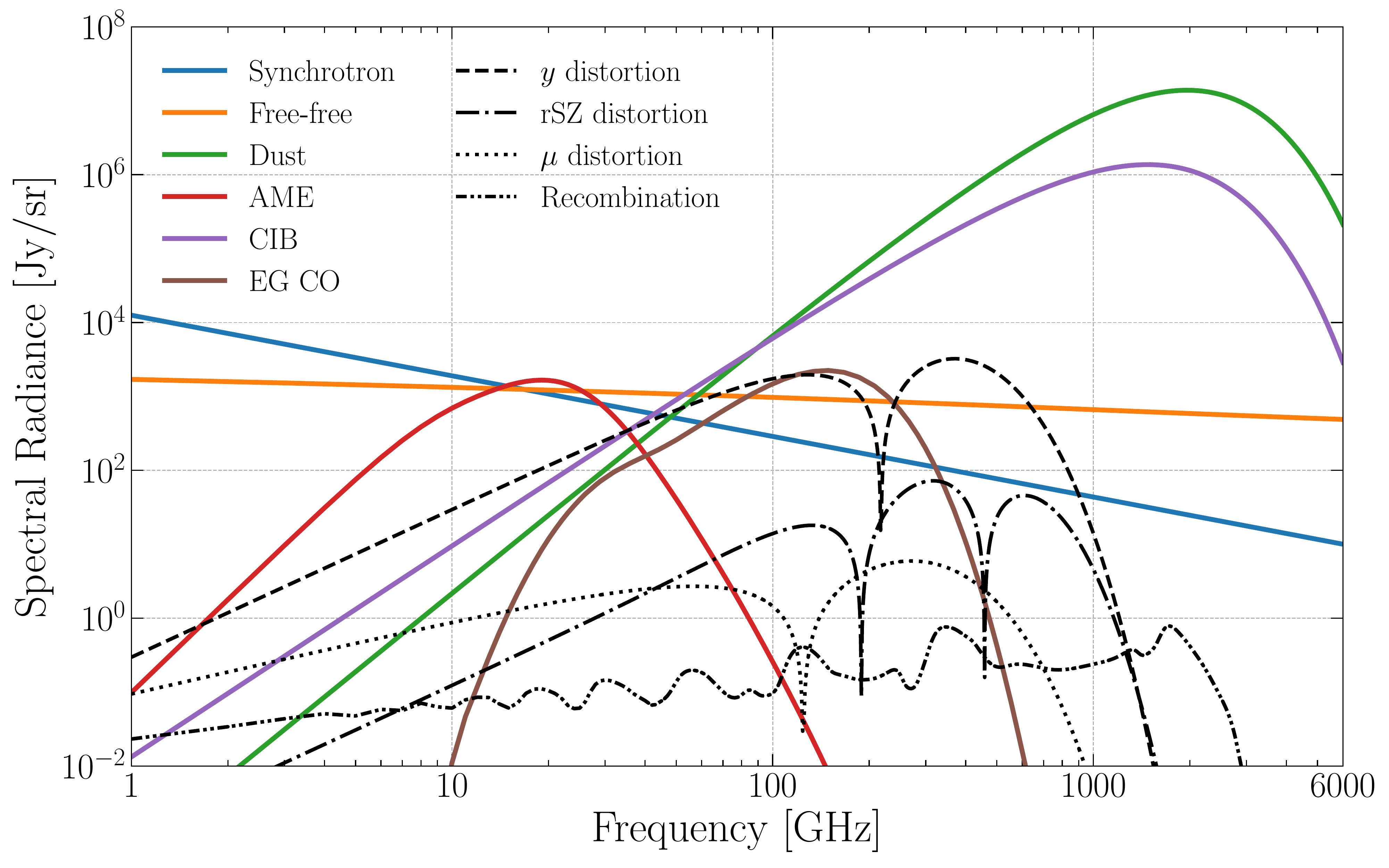}
\end{tabular}
\end{center}
\vspace{-8mm}
\caption[Foreground Overview]
{\small 
\close10
Broadband Galactic and extra-galactic foregrounds 
lie in the same frequency range as the CMB 
and its spectral distortions.
The Galactic signals include synchrotron (blue), 
free-free (orange) 
and anomalous microwave emission (AME, red) 
at low-frequencies 
and thermal dust emission (green) at high-frequencies. 
The cosmic infrared background (CIB, purple) 
and extra-galactic carbon monoxide 
integrated over redshift (EG CO, brown) 
also contribute at mid- to high-frequencies. 
Zodiacal emission and Galactic molecular lines 
have been excluded assuming the use of spatial information. 
The predicted spectral distortion signals are shown in black.
\label{foreground_overview}}
\end{figure}

\vspace{3mm}
\normalsize
\single12
\noindent
A number of foregrounds contribute appreciable signals
at frequencies relevant to CMB spectral distortions.
Figure~\ref{foreground_overview} presents an overview.
Synchrotron emission,
free-free emission,
and so-called anomalous microwave emission
are the dominant signals at frequencies below $\sim$70 GHz~\cite{wmap2013}.
Thermal dust emission from the diffuse interstellar cirrus
and
the zodiacal dust cloud
dominates at high frequencies~\cite{planck2016fgs}.
Additional foregrounds
from the integrated contribution 
of dust and CO line emission
in external galaxies
contribute at intermediate frequencies~\cite{planck2014CIB, mashian2016CO}.
The combined foreground signal
is 2--3 orders of magnitude brighter
than CMB spectral distortions
and must be subtracted to corresponding accuracy~\cite{fixsen2009, fixsen/etal:2011}.

A  number of techniques can be deployed to 
identify, model, and subtract foreground emission
\cite{fixsen/etal:1996, 
fixsen2009, 
fixsen/etal:2011, 
planck2014CIB, 
planck2016fgs, 
abitbol2017}.
Parametric models
fit multi-frequency data 
along individual lines of sight
to determine
parameters specifying the amplitude and frequency dependence
of each component.
More complex models
employ additional information
from spatial and frequency correlations present in the data
to isolate 
CMB distortions
from foreground emission.
Ancillary data may also be used to constrain
foreground emission,
restricting either the spatial distribution
or frequency dependence of individual foreground components.
Here we use multi-frequency parametric modeling
to quantify the impact of foreground emission on sensitivity
and identify where additional foreground measurements
would be useful\cite{abitbol2017}.
Since we ignore spatial/frequency correlations in the data,
the analysis effectively represents a worst-case scenario.
The key results of foreground analysis are:

\vspace{2mm}
\noindent
{$\bullet~$}
{\bf Astrophysical foregrounds,
not raw sensitivity,
are the limiting factor for spectral distortions.}
Although the spectral shape of CMB distortions
can be calculated to high precision,
the frequency dependence of the various foreground components
is not known to similar precision
and must be determined from the data.
If no prior constraints are placed on foreground spectral dependences,
foreground modeling degrades the sensitivity to
CMB spectral distortions
by a factor of 30
compared to the ideal case with no foreground emission.
Constraining foregrounds through external priors
(e.g. 1\% constraints on power-law emission parameters)
produces modestly better results. 
Fully subtracting the foregrounds
through parametric models
requires constraints at the $10^{-4}$ level, 
either through prior knowledge 
or from a fit of the multifrequency observations of spectrometer data.

\vspace{2mm}
\noindent
{$\bullet~$}
{\bf Data at frequencies below 100 GHz
are important to break foreground degeneracies.}
Emission from the diffuse dust cirrus
and the cosmic infrared background
dominates the sky at frequencies above 600 GHz,
beyond the Wien cutoff in the CMB spectrum, 
and are the main foreground contaminants above 100 GHz, 
where CMB spectral distortions are largest. 
Data in many frequency channels above 100 GHz, 
all the way into the THz range, 
can readily obtain high signal-to-noise measurements
on those high-frequency foreground components 
with sufficient redundancy to validate foreground emission models 
or refine them if necessary. 
At lower frequencies, 
confusion among multiple foreground components
(synchrotron, free-free, AME, and extragalactic line emission)
overlap with the CMB
and require high signal-to-noise ratio data
in multiple channels
to separate the CMB from the combined foregrounds.

\vspace{2mm}
\noindent
Foreground subtraction methods are rapidly evolving.
In addition to multi-frequency parametric models, 
methods exploiting the spatial structure of foreground components and/or external data sets
provide additional handles 
to separate astrophysical foregrounds from spectral distortion signals.
Note that the various foreground components
need not be identified and fit individually;
rather,
CMB spectral distortion science requires
only separation of the cosmological signal
from the combined foreground emission.
Non-parametric models or moment methods
\cite{ChlubaFM}
are promising and important avenues for continued research.

Additional insight into foreground emission
can come from dedicated foreground measurements. 
The large lever arm
between the foreground and CMB distortion amplitudes
at both high and low frequencies
allows constraints on the foreground spectral energy distribution
even at sensitivities
unable to directly probe CMB distortions.
Measurements at sub-mm wavelengths
from balloon platforms
can constrain dust properties
while
data from ground-based or balloon instruments
help distinguish the competing low-frequency foregrounds.

\clearpage
\large
\noindent
{\bf 4. Mission Concept}

\vspace{2mm}
\normalsize
\single12
\noindent
Order-of-magnitude improvements over current upper limits
require continuous spectra
at modest spectral resolution,
covering 6 or more octaves in frequency
with part-per-million channel-to-channel calibration stability.
Such broad frequency coverage precludes 
ground-based measurements,
which are limited to the available atmospheric windows
at frequencies below 300 GHz.
Balloon missions can play an important role as technology pathfinders,
but are limited in integration time
and environmental stability.
Broad frequency coverage with precision calibration
requires a space mission.

\begin{figure}[t]
\vspace{-5mm}
\begin{center}
\begin{tabular}{c}
\includegraphics[height=2.8in]{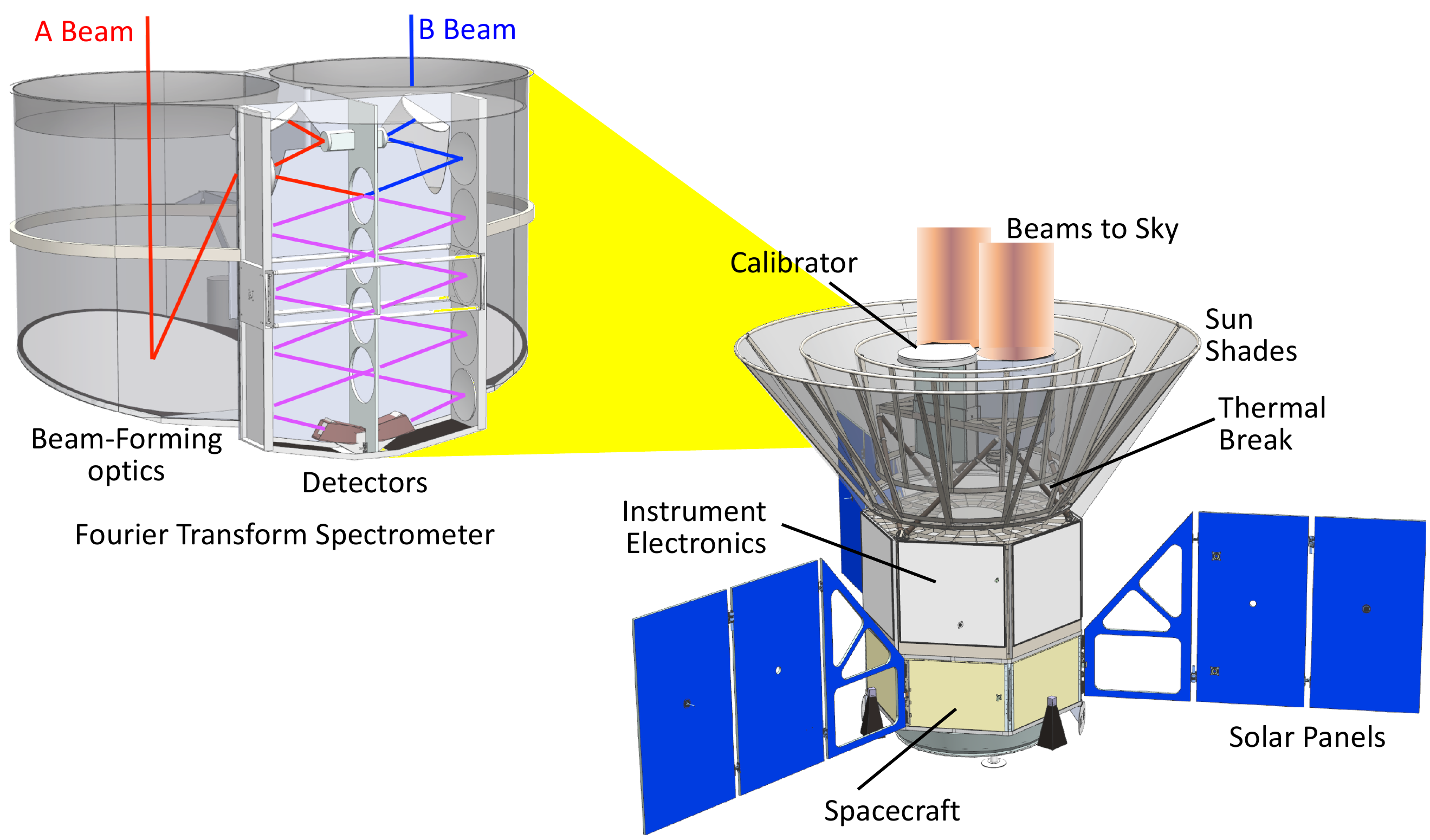}
\end{tabular}
\end{center}
\vspace{-5mm}
\caption[PIXIE mission concept]
{\small 
\close10
The PIXIE mission concept provides a costed example
for a pathfinder spectral distortion mission.
\label{pixie_mission}}
\vspace{-5mm}
\end{figure}

A broad-band Fourier transform spectrometer (FTS)
to measure CMB spectral distortions has been proposed
for several recent opportunities
(e.g.
PIXIE\cite{kogut/etal:2011, kogut/etal:2016}
as a NASA MIDEX mission, as one of the instruments on
PRISM\cite{prism:2013, prism:2014}, an ESA L-class mission,
and PRISTINE 
as an ESA F-class mission).
Figure~\ref{pixie_mission} shows the PIXIE mission.
It consists of a single cryogenic FTS
with a blackbody calibrator 
capable of moving to block either aperture.
A composite hexapod structure 
provides mechanical support and thermal isolation for the instrument.
Nested thermal shields 
provide passive cooling at 150 K
while shielding the instrument against thermal emission
from the Sun and warm spacecraft. 
A mechanical cryocooler
provides cooling from 280~K to 4.5 K, 
with intermediate stages intercepting heat from the hexapod supports
at 68 and 17 K. 
Adiabatic demagnetization refrigerators (ADRs)
cool the instrument and detectors. 
A spacectraft bus provides
power, avionics,
communication,
and propulsion.

The PIXIE observatory would be placed into a Sun-Earth L2 halo orbit
and would observe for a projected 4-year mission.
Its projected sensitivity
would detect the expected $y$-distortion
(electron pressure)
from the growth of structure
at 450 standard deviations 
and the relativistic correction (electron temperature)
at 15$\sigma$
to precisely determine the amplitude of baryonic feedback
in structure formation.
While PIXIE's raw sensitivity in principle enables a
few-standard-deviation detection of the $\mu$-distortion
from dissipation of primordial density perturbations,
astrophysical foregrounds 
are likely to prevent such a detection
($\S$3).

\begin{figure}[t]
\begin{center}
\begin{tabular}{c}
\includegraphics[height=2.3in]{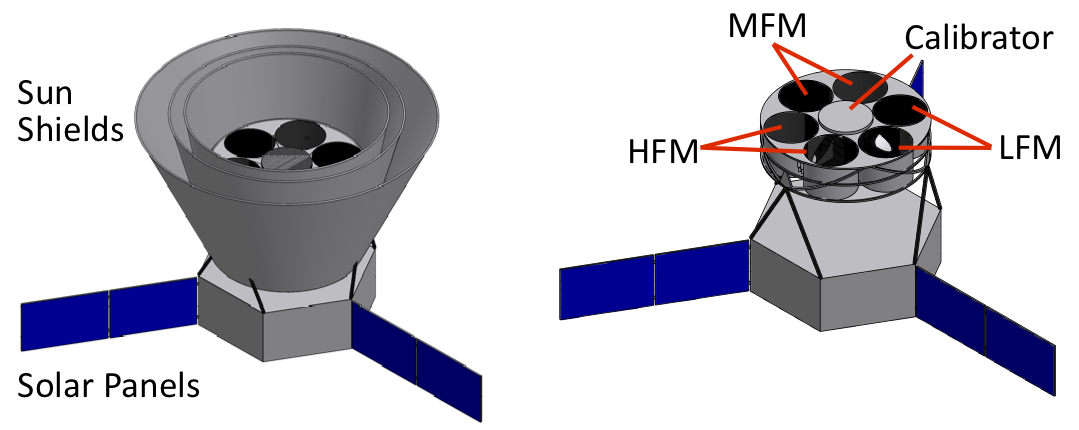}
\end{tabular}
\end{center}
\caption[Multi-module mission concept]
{\small 
\close10
Multiple FTS modules on a common spacecraft bus 
can be optimized to separate CMB spectral distortions
from competing foreground emission.
Here 3 modules are shown, although other configurations are possible.
The right panel suppresses the sun shields for clarity.
\label{mission_schematic}}
\end{figure}

We outline below a mission concept (Figure~\ref{mission_schematic}),
based on PIXIE and similar recent concepts,
capable of detecting the minimal $\mu$-distortion
from dissipation of primordial anisotropy
even for the worst-case foreground scenario.
It consists of several nearly-identical modules,
each of which uses a polarizing FTS to measure the signal difference
between the sky and a blackbody calibrator.
As with PIXIE or PRISM,
a layered combination of passive (radiative) cooling,
mechanical cryocoolers,
and sub-K coolers
maintains the FTS at a temperature of 2.725 K
(isothermal with the CMB)
and the detectors at 0.1 K.

The optical passband and maximum phase delay
differ for each module
so that the white noise level
and synthesized frequency bands
are optimized for either the CMB distortion signals
or the competing foreground emission.
A mid-frequency module (MFM) 
has etendue 4 cm$^2$ sr,
optical passband 20--600 GHz,
and 20 GHz channel width
to obtain spectral sensitivity
$\delta I_{MFM} = 1.2 \times 10^{-22}~
{\rm W~m}^{-2}~{\rm Hz}^{-1}~{\rm sr}^{-1}$
with a 1-second integration.
If foregrounds were negligible,
the MFM alone could reach the full science goals
with a 4-year integration.

Foreground subtraction requires additional sensitivity
at both higher and lower frequencies.
A high-frequency module (HFM)
with etendue
4 cm$^2$ sr,
optical passband 400--6000 GHz,
and 60 GHz channel width
has sensitivity
$\delta I_{HFM} = 6.5 \times 10^{-23}~
{\rm W~m}^{-2}~{\rm Hz}^{-1}~{\rm sr}^{-1}$
to characterize the bright high-frequency foregrounds.
The optical passband 
minimizes photon noise from the CMB
while still allowing sufficient overlap 
with the MFM to cross-calibrate the two modules.
Finally, a low-frequency module (LFM)
uses etendue
14 cm$^2$~sr,
optical passband 10--40 GHz,
and 2.5 GHz channel width
to obtain sensitivity
$\delta I_{LFM} = 2.9 \times 10^{-23}~
{\rm W~m}^{-2}~{\rm Hz}^{-1}~{\rm sr}^{-1}$
for low-frequency foregrounds.
The larger etendue prevents
signal attenuation at frequencies below the waveguide cutoff,
allowing multi-mode operation down to 10 GHz.
By cutting off the optical response at 40 GHz,
the LFM again excludes most of the CMB photon noise.
Since the LFM optical passband only covers 2 octaves,
signal dispersion over the larger phase delay
is readily controlled\footnote{
By comparison, 
controlling signal dispersion
across the MFM's 6 octaves
but with 2.5 GHz channel width
would require increasing 
the MFM transfer mirror diameter by a factor of 64.}.

\begin{figure}[b]
\begin{center}
\begin{tabular}{c}
\includegraphics[height=3.2in]{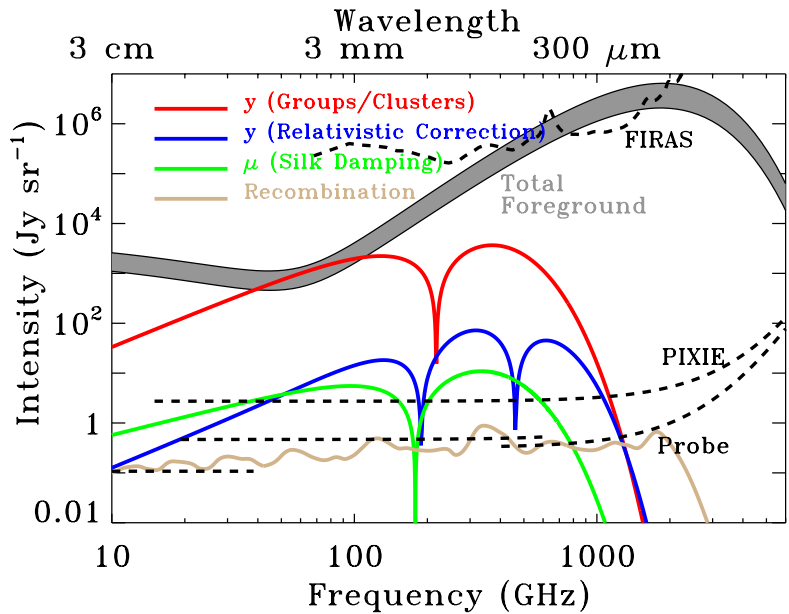}
\end{tabular}
\end{center}
\vspace{-8mm}
\caption[Mission Sensitivity]
{\small 
\close10
Spectral distortions are observable using current technology
even for worst-case foregrounds.
\label{concept_performance}}
\end{figure}

Figure~\ref{concept_performance} 
shows the predicted performance.
A mission with a single module of each type achieves
$|y| < 6.6 \times 10^{-9}$
and
$|\mu| < 5.2 \times 10^{-8}$ (95\% CL)
within a 4-year mission.
A mission with additional modules
(4 LFM, 4 MFM, and 1 HFM)
achieves limits
$|y| < 3.3\times 10^{-9}$, 
$|\mu| < 1.9\times 10^{-8}$ (95\% CL), 
and could detect the primordial hydrogen and helium recombination lines 
at $2\sigma$
within a 10-year mission
even for pessimistic foreground assumptions.

Incremental progress is possible, even likely.
A ``full'' mission ($4 \times 4 \times 1$) with no foregrounds 
detects $\mu$ at 77$\sigma$ 
(compared to 2.1$\sigma$ with foregrounds).
Even a minimal mission such as PIXIE
can detect the $y$ distortion and relativistic correction at high significance,
while opening a wide discovery space 
for new physics through the $\mu$ distortion.
Pathfinders,
either from space or balloon platforms,
can provide new foreground measurements,
directly constraining foreground emission
while determining the extent to which 
spatial/frequency correlations
can improve foreground subtraction.

Each module is based entirely on existing technologies;
no technology development is required.
While no formal cost estimates are available for this concept,
full missions using a single instrument module
have been proposed.
The PIXIE mission has been proposed to NASA's MIDEX  program
with cost cap \$250M (FY17)
while the PRISM mission
(including both a large imager and a smaller spectrometer)
has been proposed to ESA's L-class program.
Instrument costs are dominated by 
the cryogenic cooling,
which is common to all modules.

\vspace{5mm}
\large
\noindent
{\bf 5. Conclusions}

\vspace{3mm}
\normalsize
\single12
\noindent
Simple upgrades to the seminal FIRAS instrument
would improve sensitivity by three orders of magnitude or more,
providing new tests for the standard cosmology
while opening new windows for discovery.
No new technologies are required;
both the detectors,
cryogenics,
and instrumentation 
have been demonstrated.

Detection of CMB spectral distortions
is primarily limited by the need to identify and subtract
astrophysical foregrounds.
As such, new foreground data
combined with sustained effort developing methods to
identify, model, and subtract foreground emission
could significantly reduce the instrument noise levels
required to reach specific science goals.
Even in a worst-case foreground scenario,
much of the science goals could be captured
by a single FTS
within the cost caps of the NASA MIDEX program.
A more ambitious mission
using multiple copies of a basic FTS design
could reach a fundamental sensitivity threshold
to detect the distortions 
from primordial density perturbations,
providing an independent test of inflation
on physical scales orders of magnitude beyond any other measurement.

\clearpage



\end{document}